# Measuring accurately liquid and tissue surface tension with a compression plate tensiometer

*Abbas Mgharbel, Hélène Delanoë-Ayari and Jean-Paul Rieu*

Université de Lyon, F-69000, France; Univ. Lyon 1, Laboratoire PMCN; CNRS, UMR 5586; F-69622 Villeurbanne Cedex

Correspondence to JPR: rieu@lpmcn.univ-lyon1.fr

Final character count: ~36 500

**Abstract**

Apparent tissue surface tension allows the quantification of cell-cell cohesion and was reported to be a powerful indicator for the cellular rearrangements that take place during embryonic development or for cancer progression. The measurement is realized with a parallel compression plate tensiometer using the capillary laws. Although it was introduced more than a decade ago, it is based on various geometrical or physical approximations. Surprisingly, these approximations have never been tested. Using a novel tensiometer, we compare the two currently used methods to measure tissue surface tension and propose a third one, based on a local polynomial fit (LPF) of the profile of compressed droplets or cell aggregates. We show the importance of measuring the contact angle between the plate and the drop/aggregate to obtain real accurate measurement of surface tension when applying existing methods. We can suspect that many reported values of surface tension are greatly affected because of not handling this parameter properly. We show then the benefit of using the newly introduced LPF method, which is not dependent on this parameter. These findings are confirmed by generating numerically compressed droplet profiles and testing the robustness and the sensitivity to errors of the different methods.

**INTRODUCTION**

It is now well established that on a time scale of hours, embryonic tissues mimic the behavior of highly viscous liquid droplets (Beysens D et al, 2000). In the absence of external forces, irregular tissue fragments or aggregates of reaggregated cells round up into spherical shapes and fuse when they are brought into mutual contact (Gordon R et al, 1970). The engulfment of one tissue type by another via spreading, and the sorting of cell types in heterotypic mixtures are other examples of liquid-like behaviors (Steinberg MS, 1962; Steinberg MS, 1970; Technau U & Holstein TW, 1992). The very same final tissue configuration could be arrived at by an entirely different pathway (*i.e.*, sorting-out or spreading; Steinberg MS, 1963; Steinberg MS, 1970). All these processes are similar to the rounding-up, coalescence or demixing of immiscible liquids which are driven by surface tension σ and resisted by viscosity $\eta$ (Gordon R et al, 1970). Interestingly enough, these two quantities are accessible experimentally. In the case of tissues, the apparent surface tension is measured using a compression plate tensiometer (Foty RA et al, 1994) while the apparent viscosity follows from the analysis of aggregate shape relaxation kinetics (Gordon R et al, 1970; Rieu & Sawada, 2002; Mombach JCM et al, 2005).

Steinberg proposed that cell sorting is driven by surface energy minimization, arising from cellular adhesive interactions (Differential adhesion hypothesis, DAH, Steinberg MS, 1963). He concluded that mixed populations of sufficiently mobile cells rearrange so that the less cohesive cells envelop the more cohesive ones (Steinberg MS, 1970). Experimentally, measurements of apparent aggregate surface tensions have shown that a cell aggregate of lower surface tension tends to envelop one of higher surface tension to which it adheres (Foty RA et al, 1996). The link between surface tension and adhesive molecules expressed by tissues was done using L cell aggregates transfected to express N-, P- or E-cadherin in varied, measured amounts: a direct, linear correlation was observed between apparent surface tension and cadherin expression level (Foty & Steinberg MS, 2005; Hegedüs B et al, 2006). Other factors, such as cell contractility and rigidity, have been suggested to play a role in cell sorting, but have not yet been explored extensively (Harris A, 1975; Brodland G, 2002; Krieg M et al, 2008).

For biological applications, tissue surface tensiometry is a new technology to explore fundamental issues regarding cell-cell and cell-substratum interaction in (i) morphogenesis, (ii) cancer progression and (iii) tissue engineering. Foty & Steinberg (2004) and Marga F et al. (2007) recently reviewed these issues, which we briefly summarize here.

(i) The analysis of surface tension and sorting/envelopment behaviour of germ layer progenitor cells in amphibians (Davis GS et al, 1997) and zebrafish (Schötz E et al, 2008) suggests that surface tension is involved in guiding germ layer morphogenesis during gastrulation. Downregulation of E-cadherin levels in the later study leads to a decrease in the measured surface tension and a corresponding reversal of germ layer positioning in cell sorting experiments.

(ii) It is generally assumed that the invasiveness of cancer cells largely depends on a loss of cell cohesion. Cadherins have been linked with transition to malignancy for a variety of tumors. In particular, the expression of E-cadherin often, but not always, inversely correlates with tumor aggressiveness (Foty & Steinberg MS, 2004). However, for brain tumors, it was shown that dexamethasone mediated decreased invasiveness correlates with increased aggregate surface tension (*i.e.*, cohesivity) but not with N-cadherin expression (Winters BS et al, 2005). Surface tension represents a global property of a tissue that may depend from other interactions than the only cadherin-cadherin interactions. $\alpha 5\beta 1$ integrin-fibronectin interactions can indeed mediate strong cohesion (Robinson EE et al, 2004). The interactions of a cell with the surrounding matrix are also, of course very important to control the invasive properties of tumors (Hegedüs B et al, 2006).

(iii) Tissue engineering aims at reproducing morphogenesis in the laboratory, *i.e.*, to fabricate replacement organs for regenerative medicine. It has been shown that the liquid properties of some tissues, in particular the capability to fuse or reaggregate, may be used to self-assemble cellular aggregates into 3D living structures (Marga F et al, 2007). Surface tension and viscosity are parameters that can be used both experimentally and theoretically to control and predict this self- or re-assembly. In addition, surface tension can also strongly influence the ability of tissues to interact with other biomaterials (Ryan PL et al, 2001).

Currently, the only available quantitative method to measure the apparent tissue surface tension ($\sigma$) is by compression plate tensiometry (Foty RA et al, 1994; Hegedüs B et al, 2006). As cell aggregates behave as solid elastic materials at short time scales, it is also possible to follow their viscoelastic response at short time scales with this apparatus (Forgacs G et al, 1998). At long time scales, once elastic forces are relaxed, $\sigma$ is measured assuming that cell aggregates verify the same physical laws of capillarity as liquid droplets.

When a droplet is compressed between two identical plates (*i.e.*, with identical surface properties), it has a rotational symmetry around the z-axis and a reflection symmetry with respect to its equatorial plane, in which it has the two principal radii of curvatures $R_1$ and $R_2$ shown in Fig. 1A. $R_3$ is the radius of the droplet's circular area of contact with the compression plates. The compression force $F$ applied to the upper (or lower) plate is balanced by two capillary forces each proportional to $\sigma$. The first one is due to the excess pressure inside the drop due to curvature given by the Laplace formula $\Delta p = \sigma(1/R_1 + 1/R_2)$. When the radii of curvature are positive as it is generally the case with cellular aggregates, this first term is always positive (*i.e.*, repulsion between the two plates). The second term is proportional to the drop perimeter and is always negative (*i.e.*, attraction between the two plates). For an arbitrary horizontal plane, at mechanical equilibrium, the equilibrium condition when evaluated along the vertical axis, implies $F = \Delta p\, A - \sigma P \sin\phi$. Here $A$ and $P$ represent the cross-sectional area and the perimeter of the liquid drop in this plane respectively, and $\phi$ is the angle between the horizontal and the tangent to the profile of liquid drop at the plane. For the horizontal boundary plane located just underneath the upper plate, $A = \pi R_3^2$, $P = 2\pi R_3$ and $\phi = \theta$. Thus, by using the Laplace formula for $\Delta p$, one obtains:

$$F = \left(\pi R_3^2 \left(\frac{1}{R_1} + \frac{1}{R_2}\right) - 2\pi R_3 \sin\theta\right) \sigma = L_P\, \sigma \qquad (1)$$

where $L_P$ is therefore just a geometrical parameter depending on $R_1$, $R_2$, $R_3$ and $\theta$. Similarly, for the median plane of the compressed drop at $H/2$, where $H$ is the compressed drop height, one has $A = \pi R_1^2$, $P = 2\pi R_1$ and $\phi = \pi/2$. The force depends on the geometrical parameter $L_M$:

$$F = \pi R_1 \left(\frac{R_1}{R_2} - 1\right) \sigma = L_M\, \sigma \qquad (2)$$

These two equations are equivalent as the force in each droplet horizontal plane is of course conserved along vertical axis. While $R_1$ and $F$ can be accurately measured, the determination of $\sigma$ from either one of these two equations requires the measured values for $R_2$ and/or $\theta$ and $R_3$ that can only be obtained with large errors. In earlier published studies this problem has been circumvented by making the approximation that the lateral profile of the drop is a portion of circular arc (Foty RA et al, 1996; Schötz E et al, 2008):

$$R_3 = R_1 - R_2 + \sqrt{R_2^2 - (H/2)^2} \qquad (3)$$

In order to even simplify the analysis, plates are generally treated to prevent aggregate/plate adhesion and it is assumed that $\theta = 0°$, $R_2 = H/2$ and $R_3 = R_1 - R_2$ (Davis et al, 1997).

Norotte et al. (2008) have used another method based on the exact solution of Laplace equation and on the only measurements of $H$ and $R_1$. They claim their method is weakly sensitive on the angle, as long as $\theta \leq 20°$. They also claim that the previously existing method, based on the circular arc approximation (CA method), fails to give consistent results in a certain range of the compressive force or contact angle. But they did not present any quantitative analysis on the geometrical parameter sensibility of the different methods.

Although, the Exact Laplace Profile (ELP) method in principle uses exact thermodynamics and gives a rigorous estimation of surface tension in case of liquid droplets, we found that it still requires the use of experimental parameters $\theta$ and $H$ that may suffer large experimental errors. Impose $\theta \approx 0°$ may be sometimes difficult because cell aggregates may adhere to the plates after prolonged compression despite the fact that coating minimizing adhesion are used. In our experience, $\theta$ may vary during a compression experiment and $H$ is a difficult parameter to measure optically with high accuracy because of a number of interfering factors, such as; the large field of focus, the imperfect parallelism between plates, light multi-reflections or optical aberrations. In this study we used the force signal to determine the exact position at which the upper plate contacts the droplet. *H* is easily obtained from the aggregate height before compression by subtracting the motor vertical displacement done to compress the aggregate from this position and by adding the deflexion of the cantilever. For this method the resolution depends on the force signal sensitivity (0.1 µN for our study) and the Z-motor minimal step (0.04 µm here).

The motivation for the present work is to establish a more direct and robust method to measure the absolute values of apparent tissue surface tensions accurately and reliably independent of the contact angle. It is based on a direct measurement of $R_1$ and $R_2$ using a local polynomial fit (LPF) at equatorial plane and Eq. (2). In addition, we compare the three existing methods (CA, ELP and LPF methods) and evaluate their respective sensitivity to experimental errors. For that purpose we have performed compression experiments on water drops in mineral oil (W-O), air bubbles in culture medium (A-M) and embryonic cell aggregates in culture medium (C-M) with a self-made tissue tensiometer. We have also evaluated the robustness and the sensitivity to errors of the three methods by generating numerically compressed droplet profiles.

**MATERIALS AND METHODS**

**The compression plate tensiometer**

We designed and build our own surface tension apparatus: a droplet or aggregate is compressed between two parallel glass plates (fig1A). The lower compression plate consisting of a 2-mm thick borosilicate glass, is located at the bottom of a medium chamber, and is moved in the x,y,z directions through an electronic micromanipulator (MP285, Sutter Instrument). The upper compression plate, made off a cover glass, is connected through an inox wire (diameter 0.8 mm) to a copper-beryllium cantilever (spring constant $k\sim0.36$ N/m). The cantilever deflexion is measured with a non-contact eddy current position measurement apparatus (DT 3701-U1-A-C3, micro-epsilon). The complete setup is mounted in a thermally isolated chamber to maintain the desired temperature using a thermal resistance, which is controlled by a Lakeshore 331 apparatus. The aggregate profile is recorded using a binocular (MZ16, Leica) and a digital camera (A 686M , Pixelink). The lightning is adjusted by a KL 1500 LCD cold light source (Schott) through "flexible tubes". The whole setup is controlled with Labview and image analysis is performed with matlab.

The chamber in which aggregates are deposited contains an opening, to facilitate displacements and choice of aggregates. The free surface is covered with a thick mineral oil layer to prevent evaporation. Glasses surfaces are carefully cleaned with soap and pure water (sonicated 30 min with 2% Microson detergent, Fisher Bioblock France). They are first made hydrophobic by silanization with perfluorosilane (ABCR, F06179) then incubated in 10mg/ml Pluronic F-127 (Sigma) for 5 min and finally rinsed briefly with water and dried. This treatment assures a minimum of aggregates adhesiveness. Each droplet or aggregate was subjected to at least four compressions. For tissues, we waited at least 30 min between two compressions in order for the aggregate to reach shape equilibrium.

**The ELP (Exact Laplace Profile) method**

$R_1$ and θ are measured directly on the images of compressed aggregates. For the determination of *H*, it is important to use the deflexion signal to evaluate precisely the position at which the upper plate contacts the aggregate. We found that this position can be hidden on the image due to slight inclination of upper compression plate, of optical axis or of illumination. In this study, $H$ is determined with a resolution of about 2 μm. Following Norotte et al. (2008), when gravity is neglected, a dimensionless parameter α is calculated by solving numerically the following equation:

$$\frac{H}{2R_1} = f_\theta(\alpha) = \int_{\beta(\alpha,\theta)}^{1} dx \left( (\frac{x}{\alpha x^2 + 1 - \alpha})^2 - 1 \right)^{-1/2} \qquad (4)$$

where $\beta(\alpha,\theta) = \left( \sin(\theta) + \sqrt{\sin^2(\theta) + 4\alpha(\alpha-1)} \right) / 2\alpha$ and σ is then given by

$$F = 2\pi R_1 (\alpha - 1) \sigma = L_E \, \sigma \qquad (5)$$

**The LPF (Local Polynomial Fit) method**

We use Eq. (2) to calculate surface tension from the measurement of the two radii of curvature $R_1$ and $R_2$ (Fig. 1A). This requires the analysis of the aggregate profile $r(z)$ along vertical axis $z$. It is drawn by hand on the right and left side of the aggregate. We did not make any automated contour analysis because of inhomogeneities and varying lightning conditions during subsequent compression steps.

A curve can always be approximated locally by a second order polynomial $r = az^2 + bz + c$, the local radius of curvature is then given by $R_2 = \frac{1}{2a}$. The narrower the window on which the profile will be fitted, the lower the residual $dr = r_{predicted} - r_{data}$ on the fitted profile and the better the evaluation on the curvature will be (see Fig 3). Of course, we are limited by the image resolution on the contour and, in the case of aggregates, by the surface roughness. This prevents us from using two small windows. In practise, this window is adjusted between one fourth and one half of the whole side contour. $R_1$ is then taken to be half of the distance between points having the highest $r$-coordinates on both right and left fitted contours.

For each side, the error of $R_1$ is set to the maximum value of the residual $dr$ and the error on $R_2$ is given by the 95% confidence interval given on the polynomial fit using the curve fitting toolbox of matlab. The error on both sides is averaged and divided by $2^{1/2}$. It is about 3 μm and 10μm for $R_1$ and $R_2$ respectively.

**The CA (Circular Arc) method**

The complete contour on both sides of the aggregate is approximated by a circular arc as done in previous reported studies except for Norotte et al. (2008). $R_2$ is obtained by a fit of the previously recorded profile (see LPF method section) using the function fitellipse.m under matlab (http://www.mathworks.com/matlabcentral/fileexchange/15125). We have calculated the surface tension using this circular arc approximation in three different ways: (i) using Equation (2), where the force is expressed at median plane (CAcm method), (ii) by combining equations (1) and (3), where force is expressed on plates, with θ=0° (original CA method), and (iii) by combining equations (1) and (3) using the measured θ (CAcp). $R_1$ is obtained in the same way for ELP, LPF and CA methods. When it is needed, *i.e.*, for the CA and CAcp methods, $H$ is obtained as explained in the ELP method section.

**Aggregate formation**

Mouse embryonal carcinoma F9 cell line was a generous gift from S. Nagafuchi (Nagafuchi A et al, 1987). Cells were maintained in DMEM (41965-039, GIBCO) supplemented with 10% foetal bovine serum (2902 P-232310, Biotech GmbH). For the aggregate formation, cells were dissociated and reassemble in 25 µl hanging drops containing between 1000 and 8000 cells (Robinson et al, 2003). After two days, the newly formed aggregates were transferred to 24 well plates containing fresh medium and then put on a gyratory shaker for two more days. For the compression measurements, the cell aggregates were transferred to $CO_2$ independent medium (18045-054, GIBCO).

**RESULTS**

We first compressed air bubbles in culture medium ($CO_2$ independent medium, 37°C). These bubbles are easily nucleated by pouring cold medium on the tensiometer plates already at 37°C and waiting 2 hours for temperature stabilization. The complementary contact angle θ is small when glass is clean and hydrophilic. But because glass surfaces are frequently re-used, θ may change from one experiment to the other, or after successive compressions. Only for half of the compressions investigated (n=20 compression steps corresponding to 3 different bubbles), we could really obtained an angle $\theta \leq 5°$ for all four recorded glass/air/medium contact lines (Fig. 1B). The angle was comprised between 10° and 25° otherwise. Fig. 2A shows the plot of geometrical parameters $L_P$, $L_M$ or $L_E$ defined in Eqs. (1), (2) and (5) for each method as a function of the deflection transmitted to the upper plate (δ=F/k where $F$ is the force). The points in the graph correspond to three different bubbles. This plot displays a linear relationship between δ and L for each method, as expected from capillary laws and the slope k/σ gives the surface tension σ. However, with ELP and CA methods, points are much dispersed and the surface tension values (σ=56±2.6 and 39±3.1 mN/m respectively) are significantly different from a direct measurement using the wilhelmy plate pressure sensor of a Langmuir trough (NIMA, England): σ=51±2 mN/m. In contrast, the LPF method provides the correct value, with the lower error: σ=51.2±1.2 mN/m (errors Δσ are calculated from the 95% confidence interval on fitted slopes by the curve fitting toolbox of matlab). When it is calculated at the median plane, the circular arc approximation (CAcm) gives also a reasonable agreement although the value is slightly lower (σ=48±1.4 mN/m)..

We followed up our studied by compression analysis of water droplet, immerged in mineral oil (Sigma) at room temperature. The complementary contact angle was approximately θ=28±5 ° for the five investigated droplets, with minimal changes from one compressions to the other. The original CA method that ignores such a large angle (*i.e.*, it uses Eqs. (1) and (3) with θ=0°) gives clearly an underestimated surface tension. The curve $L_P$ versus the deflexion is not linear. But if we use only one or two compression points as done in most reported papers except for Norotte et al. (2008), the estimated surface tension can be more than two times smaller than the expected one. It is for instance the case, if we use the point at deflexion 30 µm corresponding to nearly a 50% deformation (see dotted line in Fig. 2B). When the ELP and the CAcp (calculated on plates) are used with the angle measured on images, they give similar results to the angle independent LPF method. Values of surface tension are σ=18.5±1.5, 17.4±1 and 17.3±1 mN/m for ELP, CAcp and LPF methods respectively. This is in agreement with reported values for water/oil interfacial tension (du Nouy PL, 1925; Norotte C et al, 2008). Fig. 2C shows the angle sensibility of ELP method when angle is not properly chosen. In this case, the surface tension varies a lot, σ=12.7±0.8, 13.0±0.8, 14.0±0.7, 16.2±0.7 and 20.8±2 mN/m for θ=0°, 10°, 20°, 30° and 40° respectively, while points are still linearly aligned. The error is not increasing particularly, and therefore one has to keep in mind that a low error or a linear alignment as predicted by Laplace law is not an insurance of correct surface tension measurement, using ELP method.

Embryonic F9 cell aggregates constitute the third (biological) system we investigated. When compressed, the aggregates show a complementary contact angle between 20° and 30° (Fig. 1D). With such values, the non angle corrected CA method gives a much lower surface tension value ($\sigma$=3.3±2.1 mN/m) than the angle independent methods: $\sigma$=5.3±0.7 and 5.6±0.8 mN/m for LPF and CAcm respectively (Fig. 2D). When the angle-measurements from the image profiles are properly taken into account, ELP and CAcp methods give similar values but with higher errors: $\sigma$=5.3±1.0 and 5.0±1.0 mN/m respectively. Alternatively, the choice of the correct angle may be done by an adjustment of the compressions points of angle dependant and independent methods. This provides a way to evaluate the complementary angle when image profile is not clear because of very small ($\theta$ <10°) or because image is partially hidden by for instance the upper plate, as in Fig. 1D.

In order to validate these findings, we have generated exact droplet profiles following the study of Norotte et al. (2008). For a given volume $V$, a given separation $H$ between plates, a given contact angle and surface tension (we used here $\sigma$=5 mN/m), the shape of the droplet and the force $F$ exerted on the plates can be determined. We have recalculated the surface tension using the different methods by eventually introducing small errors in the input geometrical parameters ($\theta$, see below, and $H$, not shown) in order to simulate the 'measurement inaccuracies' of a real experiment. Of course, the ELP method gives always the exact value when we introduced the correct angle (black curves in Figs. 3B-C).

For a null complementary contact angle, one can appreciate in Fig. 3A the deviation of the real profile (green curve) to the circular arc (blue curve). The deviation is only localised in the vicinity of the plate but it results a 15% error in the determination of $R_3$ used in Eq. (1) and thus in $\sigma$. This explains why the CA method based on this equation has the largest error on the surface tension even when the complementary contact angle is 0° (Fig. 3B). When Eq. (2) is used, the error in $\sigma$ by using the CAcm or LPF method is introduced by the measurement of $R_2$ itself. Both methods underestimate $\sigma$ because $R_2$ is slightly overestimated. The deviation to the real value ($\sigma$=5 mN/m) decreases for large compressions (*i.e.*, large deformation parameter $\varepsilon$=Ho-H/Ho where Ho is the uncompressed aggregate height). The LPF method gives the best results when the proportion $H/f$ of the profile around the median plane used for the local second order polynomial fit is not too large: $\sigma$=4.88 and 4.58 mN/m for $f$=4 and 2 respectively for a large compression ($\varepsilon$=0.5). The CAcm method gives an intermediate value 4.66 mN/m when using the full profile ($f$=1).

When the complementary contact angle is larger (*i.e.*, $\theta$=20°, Fig. 3C), again the LPF method provides the best results for a narrow fitting window ($f$=4) followed by the CAcm method (Fig. 3C). The error is even lower compared to the analysis where $\theta$ = 0°: $\sigma$=4.90 and 4.78 mN/m for the LPF ($f$=4) and CAcm methods respectively for $\varepsilon$=0.5. Error is much larger with the ELP method when an error of only 5° is introduced on $\theta$: for $\theta$=25°, $\sigma$=5.5 mN/m (*i.e.*, 10% error). The CA methods using $\theta$=0° gives completely wrong results.

**DISCUSSION**

In this report, we have carefully investigated the effect of the geometrical approximations on measuring the surface tension of liquids and tissues. For tissues, we have used a cell line that shows perfect spheroids for nearly all prepared aggregates. When they are compressed, we have found they respond well as capillary droplets as their profile is well described by the Laplace profile (green curve in Fig. 1D) and they verify the capillary laws (Fig. 2D). We have also tested cell lines for which aggregates are often more irregular and present flat faces and sharp angles (CHO, chicken retina). We currently cannot assess whether an apparent surface tension is meaningful for those aggregates but this issue is out of the scope of this paper.

We have shown that when tissue surface tension measurements are properly employed, the three methods to analyse surface tension from compressed droplet (ELP, LPF and CA) give similar results with slightly different errors. Nevertheless, each method has a different sensibility to geometrical parameters making some of them more robust than others.

To date, except in the study of Norotte *et al.* (2008), the circular arc approximation method is the only one used to measure tissue surface tension. Even when the deviation to the circular arc profile is evident near the plates (Fig. 3A), the radius of curvature $R_2$ at median plane is correctly calculated. As a result, the CAcm method which consists in a circular arc fit combined with the force Eq. (2) at median plane provides very acceptable results (5-10% error on simulated drops depending on contact angle and compression rate, Fig. 3B-C). The simultaneous use of Eq. (1) with θ=0 and of Eq. (3) leads to some inconsistencies when the complementary contact angle actually deviates from zero The result is that the measured surface tension is systematically lower than the actual tissue surface tension value, and can lead to an underestimation of more than 100% (Figs. 2B,D and 3C). One can therefore postulate that some of the reported surface tension values are incorrect because of a finite contact angle. Specifically, we have identified two studies where contact angles are visibly large (Foty RA et al, 1998; Schötz E et al, 2008). In addition, it is important to note that in most of the other reported studies images of compressed aggregates are not shown.

Inserting the correct angle in Eq. (1) allows a correct estimate of σ, when using the CAcp method. However, there is no need to use Eq. (1) at the plane of the plate, because this requires the input of four geometrical parameters $R_1$, $R_2$, θ and $H$, of which especially $R_2$ and θ are usually measured with large errors. In contrast, the ELP method, based on the exact solution of Laplace equation, is more robust from a computational point of view. The resolution of Eqs. (4) and (5) is numerically easy and does not necessitate a least squared fit of the radius of curvature $R_2$ as in case of LPF or CA methods. However, we found that an error of 5° on θ introduces non negligible relative errors on the surface tension σ (about 10%). Such an error is not particularly exaggerated for cell aggregates when θ is small because cell aggregates are rough, not as regular as fluid droplets. The contact angle may also change after several subsequent compressions. The ELP method also requires the input of the aggregate compressed height $H$ which may suffer errors for instance due to optical aberrations, hidden part of the droplets.

As a conclusion, when performing compression experiments, we strongly recommend the use of an angle independent method that requires only the two parameters $R_1$ and $R_2$. The CAcm method may have the lowest error when aggregate is symmetric but rough. Otherwise, the proposed LPF method is robust and provides the lowest error when the profile is smooth as the fitting window can be narrowed. This method is a straightforward application of Laplace equation. $R_2$ is measured by a second order polynomial fit at median plane. While other methods assume reflection symmetry with respect to equatorial plane, the LPF method deals naturally with up-down asymmetries in the aggregate profile. Such asymmetries may arise either because of a slightly titled field of view, non-perfectly parallel-aligned compression plates, the effect of gravity or because of differences in adhesion affinities and contact angles with the top and bottom compression plates. Our least squared fit code running on matlab is available upon request at Helene.Ayari@lpmcn.univ-lyon1.fr.

ACKNOWLEDGMENTS

This project was funded by the ANR "Jeunes chercheuses Jeunes Chercheurs" 2005. The research team belongs to the CNRS consortium CellTiss. A. M., H.D.-A. and J.-P.R.. designed the study; A.M. performed the experiments and the analysis. H. D.-A. contributed to data analysis; and J.-P.R. wrote the manuscript. We would like to thank Pr. A. Nagafuchi (Kumamoto University, Japan) for its generous gift of F9 cell line. We thank J. Käfer, L. Bocquet, E. Charlaix and T. Biben for useful discussions and F. Graner, G. Krens for their useful comments on the manuscript.

**COMPETING INTERESTS STATEMENT**

*The authors declare no competing interests.*

# Figures:

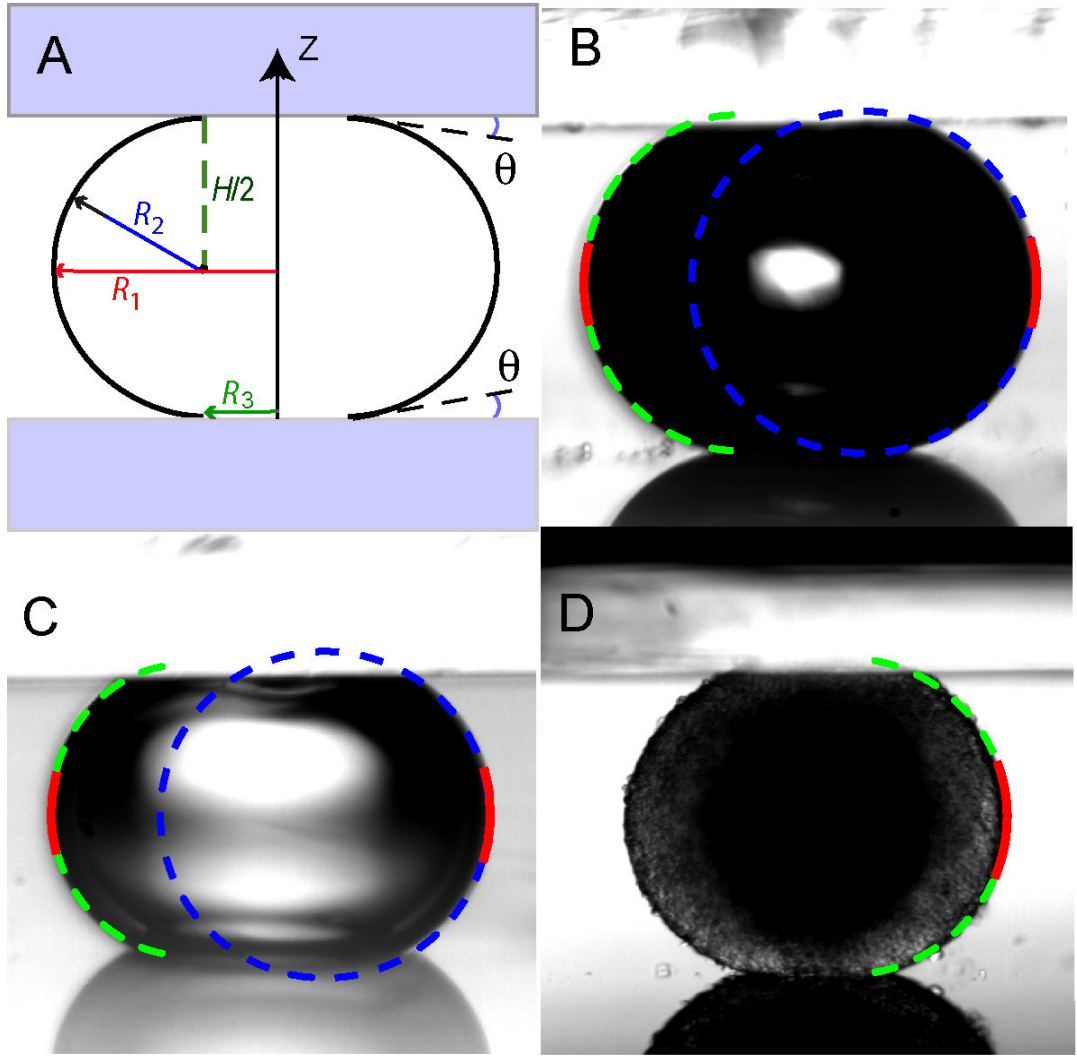

**Fig. 1.** (A) Diagram of a liquid droplet compressed between two parallel plates to which it adheres with a contact angle θ. At equilibrium, $R_1$ and $R_2$ are the two primary radii of curvature, at the droplet's equator and in a plane through its axis of symmetry, respectively. $R_3$ is the radius of the droplet's circular area of contact with either compression plate. $H$ is the distance between upper and lower compression plates. (B)-(D) Snapshots of compressed droplets: (B) Air bubble ($R_1$ =282 μm) in culture medium; (C) water droplet ($R_1$ =207μm) in mineral oil and (D) mouse embryonic cell aggregate ($R_1$ =297 μm) in culture medium. Red curves corresponds to the calculated profile with a second order polynomial fit (LPF method), dotted lines correspond to the calculated profiles with the exact Laplace profile method (ELP, green) and the circular arc approximation (blue). The latter is not represented in order to appreciate the embryonic aggregate roughness in (D).

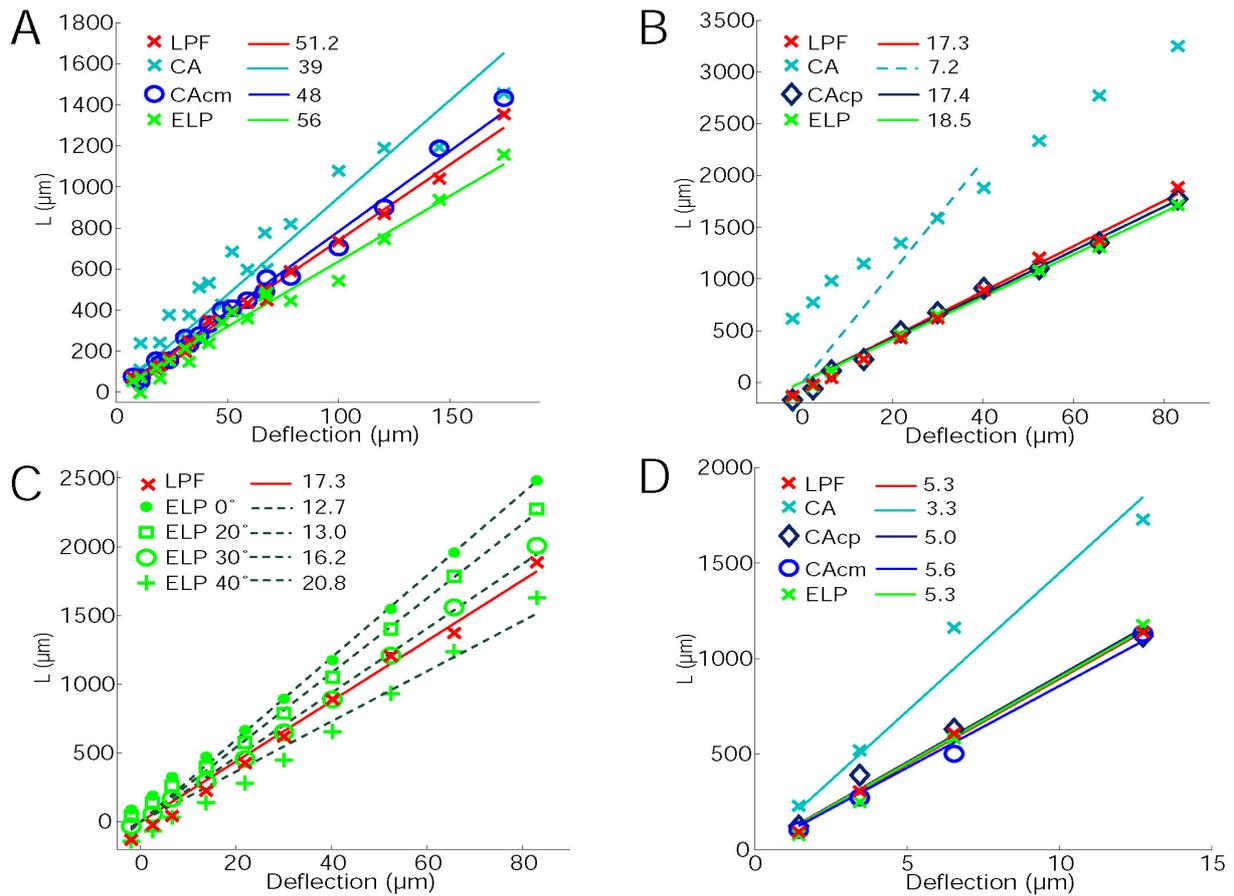

**Fig. 2:** Plots of the geometrical parameters $L_P$, $L_M$ and $L_E$ as a function of cantilever deflexion $\delta$. Surface tension values (displayed in the legend in mN/m) are obtained by taking the slope of the obtained linear fits, with $L_P$, $L_M$ and $L_E$ obtained as defined in Eqs. 1,3,5, and $\delta = F/k$ (with k~0.36 N/m for all experiments). Air bubble in $CO_2$ independent culture medium at 37°C (A), water droplet in mineral oil at room temperature (B-C) and embryonic cell aggregate in $CO_2$ independent culture medium at 37°C (D). Abbreviations of the different methods tested are as follows: ELP, Exact Laplace Profile; LPF, Local Polynomial Fit; CA: Circular Arc; CAcm, Circular Arc with force calculated at median plane; CAcp, Circular Arc with force calculated on plates.

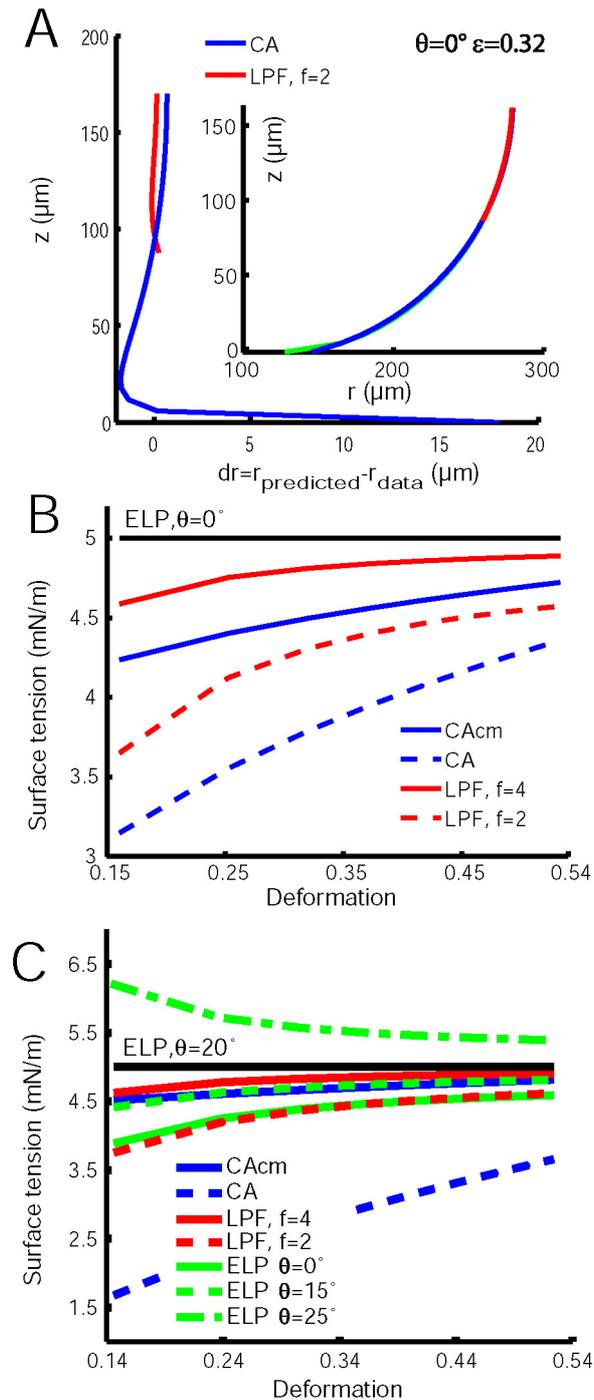

**Fig. 3:** Tests of robustness of the different methods from a numerically generated exact droplet profile following Norotte et al. (2008) with a designated surface tension 5mN/m. (A) Residual and radius (inset) of the profile estimated by the different methods for here θ=0°, a deformation parameter $\varepsilon=Ho\text{-}H/Ho$=0.32 (green, ELP method which gives the exact profile when contact angle is properly set; blue, CA methods showing a deviation near the plate; Red, profile from a polynomial fit in a central window). Surface tension evaluation as a function of the deformation parameter (B,C) for θ=0° (B), and θ=20° (C). In the later case we simulated compression and determined the surface tension with a deliberately incorrectly chosen θ, using the ELP method.